\documentclass[preprint,amsmath,amssymb,aps,10pt]{revtex4-2}

\usepackage{graphicx}
\usepackage{dcolumn}
\usepackage{bm}
\usepackage{xcolor}
\usepackage{amsfonts}
\usepackage{graphics}
\usepackage{hyperref}
\usepackage{slashed}
\usepackage{babel}
\makeatletter

\begin{document}

\title{Dual equivalence between the Maxwell-Chern-Simons theory and two self-dual massive models interacting with matter in $\mathcal{N}=2$, $d=3$ superspace}

\author{F. S. Gama}
\email{fgama@unifap.br}
\affiliation{Departamento de Ci\^{e}ncias Exatas e Tecnol\'{o}gicas, Universidade Federal do Amap\'{a}, 68903-419, Macap\'{a}, Amap\'{a}, Brazil}

\author{J. R. Nascimento}
\email{jroberto@fisica.ufpb.br}
\affiliation{Departamento de F\'{\i}sica, Universidade Federal da Para\'{\i}ba\\
 Caixa Postal 5008, 58051-970, Jo\~ao Pessoa, Para\'{\i}ba, Brazil}

\author{A. Yu. Petrov}
\email{petrov@fisica.ufpb.br}
\affiliation{Departamento de F\'{\i}sica, Universidade Federal da Para\'{\i}ba\\
 Caixa Postal 5008, 58051-970, Jo\~ao Pessoa, Para\'{\i}ba, Brazil}

\author{P. Porfirio}
\email{pporfirio@fisica.ufpb.br}
\affiliation{Departamento de F\'{\i}sica, Universidade Federal da Para\'{\i}ba\\
 Caixa Postal 5008, 58051-970, Jo\~ao Pessoa, Para\'{\i}ba, Brazil}

\begin{abstract}
We investigate the $\mathcal{N} = 2$ supersymmetric generalization of the dual equivalence between the Maxwell-Chern-Simons model and two massive self-dual models. One of the self-dual models is described by a real scalar superfield, while the other is described by a chiral spinor superfield. The equivalence is analyzed in the presence of dynamical matter chiral superfields. Initially, we establish the duality by demonstrating the existence of master actions that interpolate between the models. We then show that the field equations are identical when the proper identifications are made. Finally, we confirm the dual equivalence by defining a master generating functional, which, after changing the variables, provides the generating functionals of the Maxwell-Chern-Simons model and the two massive self-dual models coupled to matter.
\end{abstract}

\maketitle
\newpage

\section{Introduction}

The duality between two distinct field theory models is a very interesting phenomenon allowing to develop two different but equivalent physical descriptions for the same physical phenomena. Among the various examples of duality, the special role is played by the dual equivalence between the three-dimensional self-dual (SD) model and the Maxwell-Chern-Simons (MCS) one. A remarkable observation consists in the fact that while the first theory is not gauge invariant, the second is, nevertheless, there is a consistent mapping between two theories. Originally this duality has been described in \cite{TPN,DJ} for the simplest cases, that is, free SD and MCS theories. Further, this equivalence has been generalized for the presence of nontrivial couplings of vector fields (that is, the self-dual and gauge ones), and various methods to find a dual for a given extension of the SD theory, such as, for example, gauge embedding (see e.g. \cite{AINRW}) and master action (see e.g. \cite{GMS}), have been developed. Afterwards, this duality was shown to occur for supersymmetric SD and MCS theories within the superfield formulation \cite{KLRN}, which was subsequently shown to hold also in the presence of matter superfields \cite{FGNPS,FGLNPS}. Further, it has been verified for many other scenarios, including, for example, noncommutativity (see e.g. \cite{Mariz}) and Lorentz symmetry breaking scenarios (see e.g. \cite{Scarp} and references therein). We note that, in principle, such a duality, as well as any correspondence between two completely distinct theories, could have very profound implications, the paradigmatic example is the famous the holographic duality between $\mathcal{N}=4$ SYM in four dimensions and string theory in the $AdS_5\times S^5$ background known as the AdS/CFT correspondence \cite{AdsCFT,AdsCFT1,AdsCFT2} (for a review on AdfS/CFT correspondence see e.g. \cite{Hubeny} and references therein).

We will investigate the  duality between SD and MCS theories within the extended supersymmetry context. Theories with extended supersymmetry possess considerable importance due to their value in theoretical physics. Firstly, such theories demonstrate improved behavior with regard to quantum divergences in comparison to standard field theories. Indeed, the $\mathcal{N}=2$ SYM theory is finite beyond one loop, whereas the  $\mathcal{N}=4$ version is finite to all orders of perturbation theory \cite{finiteness,finiteness1,finiteness2,finiteness3}. Secondly, extended supersymmetry determines the general structure of the leading contribution to the low-energy effective action in the SYM theory, for example, in $\mathcal{N}=2$ it is described by a holomorphic function $\mathcal{F}(W)$, whereas in $\mathcal{N}=4$ it is described by a non-holomorphic function $\mathcal{H}(W,\bar W)$, where $W$ and $\bar W$ are $\mathcal{N}=2$ superfield strengths \cite{Seiberg,Seiberg1}. Lastly, theories with extended supersymmetry are significant in the context of dualities, including the already mentioned AdS/CFT correspondence, as well as the electric-magnetic duality, which is an exact symmetry of the abelian low-energy effective action of $\mathcal{N}=2$ SYM theory \cite{Electromagnetic,Electromagnetic1}.

In this paper, we formulate two SD models in $\mathcal{N}=2$, $d=3$ superspace and prove their duality with the MCS theory, first, at the tree level, second, at the level of generating functionals. As it is known, the $\mathcal{N}=2$ three-dimensional supersymmetric field theories can be formulated in the $\mathcal{N}=2$, $d=3$ superspace, which allows us to maintain the $\mathcal{N}=2$ supersymmetry manifest at all stages of the calculation \cite{N2MCS,N2MCS1,N2MCS2,N2MCS3,N2MCS4,BPS}. Earlier, various aspects of field theory models formulated in such a superspace have been studied, see e.g. \cite{Buch1,Buch2}. Therefore, a study of duality in this case is a natural problem.

The structure of the paper looks like follows. In the section II, we formulate our models in the $\mathcal{N}=2$, $d=3$ superspace and show the equivalence of the equations of motion using the master action approach. In the section III, we write down the generating functionals of these theories, and, performing the functional integrals over SD and MCS fields respectively, arrive at the some function of external currents which proves the equivalence at the quantum level.  Finally, in the section IV we summarize our results.

\section{Equivalence at the level of field equations}

As it is known (see e.g. \cite{GMS}), one of the possible methods to prove equivalence of two theories is based on defining a so-called master action including dynamical variables of both theories and defined in such a way that elimination of dynamical variables of the first theory through their equations of motion from the master action yields a classical action of the second theory, perhaps with additive terms depending only on sources, and vice versa. In our case, we are able to define even not one, but two master actions depending on the self-dual and gauge fields and corresponding sources. These master actions will be defined in this section, where we employ the definitions of $\mathcal{N}=2$ superfields introduced in \cite{Buch1}.

Explicitly, we consider the following theories. First, we introduce the Maxwell– Chern–Simons (MCS) theory defined in terms of the real scalar gauge superfield $V$,
\begin{equation}
	\label{MCS10}
	S_{MCS1}=-\frac{1}{2g^2}\int d^7z\left(G^2+mVG\right)+S_{int},.
\end{equation}
where $G=\bar{D}^{\alpha}D_{\alpha}V$ is the field strength, $S_{int}$ is the interaction term, and the action is invariant under transformations $\delta V=i\left(\bar \Lambda-\Lambda\right)$.

The corresponding self-dual model is
\begin{equation}
		\label{SD10}
		S_{SD1}=\frac{1}{2g^2}\int d^7z\left(\frac{1}{m}\sigma\bar D^\alpha D_\alpha\sigma+\sigma^2\right)+S_{int},
\end{equation}
where $\sigma$ is the scalar self-dual superfield.

Similarly, another MCS model is
\begin{equation}
		\label{MCS20}
		S_{MCS2}=-\frac{1}{2g^2}\int d^7z\left(G^2+mVG\right)+S_{int},
\end{equation} 
and the corresponding self-dual one is
\begin{eqnarray}
		\label{SD20}
		S_{SD2}&=&-\frac{1}{4mg^2}\bigg\{\int d^7z\bar \pi^\alpha\pi_\alpha+\nonumber\\ &+&
		\int d^5z\left[\frac{1}{2}\pi^\alpha i{\partial_\alpha}^\beta\pi_\beta+m\pi^\alpha\pi_\alpha\right]
		+\int d^5\bar{z}\left[\frac{1}{2}\bar \pi^\alpha i{\partial_\alpha}^\beta\bar \pi_\beta+m\bar\pi^\alpha\bar\pi_\alpha\right]\bigg\}+S_{int}.
\end{eqnarray}
Within this paper, we will explicitly define couplings of these models to currents and establish duality relations between these currents.

The first action is a functional $S_{M1}$ of the set of real scalar superfields $V$ and $\sigma$ defined in the $\mathcal{N}=2$, $d=3$ superspace, where $V$ is the  MCS superfield and $\sigma$ is the scalar self-dual superfield. In order to account for the interaction of these superfields with dynamical chiral (antichiral) matter superfields $\Phi$ and $\bar {\Phi}$, i.e. those ones satisfying the equations $D_{\alpha}\bar{\Phi}=0$, $\bar{D}_{\alpha}\Phi=0$ respectively, our first master action $S_{M_1}$ must also be a functional of $\Phi$ and $\bar \Phi$. On the assumption that such interaction occurs through linear couplings with $V$ and $\sigma$, we define $S_{M_1}$ as
\begin{equation}
	\label{master_1}
	S_{M1}=\frac{1}{g^2}\int d^7z\left(-\sigma G+\frac{\sigma^2}{2}-\frac{m}{2}VG\right)+\int d^7z\left(\sigma k+VJ\right)+S[\bar\Phi,\Phi],
\end{equation}
where $m$ is represents the mass parameter, while $g$ represents the coupling constant with mass dimension $[g]=\frac{1}{2}$. The functional $S[\bar\Phi,\Phi]$ denotes the matter sector of the master action and determines the dynamics of the chiral matter superfields. As we already noted, the gauge-invariant scalar superfield strength $G$ is given by $G=\bar D^\alpha D_\alpha V$. The currents $k$ and $J$ are constructed solely from the dynamical chiral matter superfields. With respect to the local $U(1)$ invariance of (\ref{master_1}), two conditions must be satisfied: (i) that only the gauge superfield $V$ undergoes a non-trivial gauge transformation, specifically $\delta V=i\left(\bar \Lambda-\Lambda\right)$, and (ii) that $J$ is a linear superfield satisfying the condition $J=\Pi_\frac{1}{2}J$ (or equivalently $\Pi_0 J=0$), where $\Pi_{\frac{1}{2}}:=-\Box^{-1}D^\alpha\bar D^2D_\alpha$ and $\Pi_0:=\Box^{-1}\{D^2,\bar D^2\}$ are the usual projection operators \cite{GGRS}. We note that here and further, the operator $\Box^{-1}$ does not produce any nonlocality being just an ingredient of some projecting operators since, acting on physical parts of fields and currents, all nonlocality is cancelled out. Lastly, the action $S_{M1}$ is the $\mathcal{N}=2$, $d=3$ superspace generalization of those ones defined in \cite{GMS} and \cite{FGLNPS}.

Varying (\ref{master_1}) with respect to $\sigma$  yields the following algebraic field equation
\begin{equation}
	\label{identification1}
	\sigma=G-g^2k.
\end{equation}
It should be noted that by setting the current $k=0$, the self-duality condition $\sigma= G$ is obtained, which corresponds to the $\mathcal N = 2$ version of the self-duality condition presented in Ref. \cite{TPN}. The Eq. (\ref{identification1}) allows us to eliminate  $\sigma$  from the action (\ref{master_1}). The resulting action is given by
\begin{equation}
	\label{MCS1}
	S_{MCS1}=-\frac{1}{2g^2}\int d^7z\left(G^2+mVG\right)+\int d^7z\left(VJ+Gk-\frac{g^2}{2}k^2\right)+S[\bar\Phi,\Phi].
\end{equation}
This is precisely the $\mathcal{N} = 2$ supersymmetric generalized MCS model \cite{N2MCS}, but the gauge superfield is coupled to matter in two manners, explicitly, via a minimal coupling and a non-minimal magnetic-like coupling. In addition to that, there is a Thirring-like current-current interaction. We note that this model is a natural extension of the free $\mathcal{N} = 2$ model (\ref{MCS10}). It should be noted that the gauge invariance of $S_{M1}$ under the transformations $\delta V=i\left(\bar \Lambda-\Lambda\right)$ is preserved in $S_{MCS1}$. This is expected, since the calculations above demonstrate that Eq. (\ref{master_1}) is a first-order formulation of (\ref{MCS1}), in which $V$ is the usual gauge superfield and $\sigma$ is an auxiliary superfield.

On the other hand, let us vary (\ref{master_1}) with respect to $V$. This results in the following field equation
\begin{equation}
	\label{G}
	G=\frac{1}{m}\left(g^2J-\bar D^\alpha D_\alpha\sigma\right).
\end{equation}
The identity $(\bar D^\alpha D_\alpha)^2=\Box\Pi_{\frac{1}{2}}$ permit us to rewrite this equation as
\begin{equation}
	\label{PIV}
	\Pi_{\frac{1}{2}}V=\frac{1}{m}\left(g^2\frac{\bar D^\alpha D_\alpha}{\Box}J-\Pi_{\frac{1}{2}}\sigma\right).
\end{equation}
Since $G=\Pi_\frac{1}{2}G$ and $J=\Pi_\frac{1}{2}J$, we have
\begin{equation}
	\label{identity}
	\int d^7zVG= \int d^7z\left(\Pi_{\frac{1}{2}}V\right)G \ \ \text{and} \ \ \int d^7zVJ= \int d^7z\left(\Pi_{\frac{1}{2}}V\right)J. 
\end{equation}
This means that the results (\ref{G}) and (\ref{PIV}) can be inserted into (\ref{master_1}) to derive the $\mathcal{N} = 2$ supersymmetric generalized scalar self-dual action coupled to matter:
\begin{equation}
	\begin{split}
		\label{SD1}
		S_{SD1}=&\frac{1}{2g^2}\int d^7z\left(\frac{1}{m}\sigma\bar D^\alpha D_\alpha\sigma+\sigma^2\right)+\int d^7z\left[\sigma\left(k-\frac{1}{m}J\right)+\frac{g^2}{2m}J\frac{\bar D^\alpha D_\alpha}{\Box} J\right]\\
		&+S[\bar\Phi,\Phi].
	\end{split}
\end{equation}
This result is consistent with that obtained in \cite{GMS,FGLNPS,AINRW,FGNPS}. Besides, it is  a natural extension of the free self-dual model (\ref{SD10}). As in \cite{FGLNPS}, we see that the scalar self-dual superfield is minimally coupled to $k$, but it is non-minimally coupled to $J$. Additionally, a non-local Thirring-like interaction emerges, which is common within constructing of dual theories. The dual equivalence between $S_{MCS1}$ and $S_{SD1}$ is established through observation of the fact that (\ref{MCS1}) and (\ref{SD1}) originate from the same master action $S_{M1}$ as defined in (\ref{master_1}). This duality symmetry between the models exchanges the minimal coupling $VJ$ in (\ref{MCS1}) for the non-minimal couplings in (\ref{SD1}), and  the  minimal coupling $\sigma k$ in (\ref{SD1}) for the non-minimal couplings in (\ref{MCS1}).

The dualization between $S_{MCS1}$ and $S_{SD1}$ is also revealed through an investigation of their respective field equations. In order to accomplish this, it is necessary to determine the field equations by varying both (\ref{MCS1}) with respect to $V$ and (\ref{SD1}) with respect to $\sigma$. The result of this process is 
\begin{align}
	\label{G_eq}
	\bar D^\alpha D_\alpha G+mG&=g^2J+g^2\bar D^\alpha D_\alpha k;\\
	\label{Sigma_eq}
	\bar D^\alpha D_\alpha\sigma+m\sigma&=g^2J-mg^2 k.
\end{align}
A comparison of these equations reveals that they are identical when identification (\ref{identification1}) is carried out. This would confirm the equivalence of the models at the level of field equations if the currents were not functions of dynamical chiral matter superfields. However, they depend on these superfields. Consequently, to confirm the dual equivalence, it is also necessary to demonstrate that the field equations for the matter superfields derived from $S_{MCS1}$ and $S_{SD1}$ are identical. To achieve this, we begin by varying (\ref{SD1}) with respect to $\Phi$, resulting in
\begin{equation}
	\label{matter_1}
	\frac{\delta S_{SD1}}{\delta\Phi}=0\implies\frac{\delta S}{\delta\Phi}=-\int d^7z\left[\sigma\frac{\delta k}{\delta\Phi}+\left(\frac{g^2}{m}\frac{\bar D^\alpha D_\alpha}{\Box}J-\frac{1}{m}\sigma\right)\frac{\delta J}{\delta\Phi}\right].
\end{equation}
The next step is to solve Eq. (\ref{Sigma_eq}) for $\sigma$ and substitute the result into Eq. (\ref{matter_1}). To do this, we first apply the operators $D^2$ and $\bar D^2$ to both sides of Eq. (\ref{Sigma_eq}), which gives us the constraints
\begin{equation}
	D^2\sigma=-g^2D^2k \ \ ; \ \ \bar D^2\sigma=-g^2\bar D^2k.
\end{equation} 
By applying $\bar D^\alpha D_\alpha$ to both sides of (\ref{Sigma_eq}) and making use of the constraints presented above, we obtain an inhomogeneous Klein-Gordon equation
\begin{equation}
	\label{wave_sigma}
	\left(\Box-m^2\right)\sigma=g^2\left(-mJ+\bar D^\alpha D_\alpha J+m^2k-m\bar D^\alpha D_\alpha k-\{D^2,\bar D^2\}k\right).
\end{equation}
Since the operator $\hat{\mathcal{O}}^{-1}\equiv\left(\Box-m^2\right)$ is well-defined, it is possible to express $\sigma$ solely in terms of current terms. Thus, by solving (\ref{wave_sigma}) for $\sigma$ and substituting the result into (\ref{matter_1}),  we obtain the following field equation
\begin{equation}
	\label{final_matter_eq1}
	\begin{split}
		\frac{\delta S}{\delta\Phi}=&g^2\int d^7z\bigg[\hat{\mathcal{O}}\left(mJ-\bar D^\alpha D_\alpha J-m^2k+m\bar D^\alpha D_\alpha k+\{D^2,\bar D^2\}k\right)\frac{\delta k}{\delta\Phi}\\
		&+\left(-\bar D^\alpha D_\alpha k+m\Pi_{\frac{1}{2}}k-J+\frac{m}{\Box}\bar D^\alpha D_\alpha J\right)\frac{\delta J}{\delta\Phi}\bigg].
	\end{split}
\end{equation}
Let us now find the field equation resulting from the variation of the action (\ref{MCS1}) with respect to $\Phi$. It is given by
\begin{equation}
	\label{matter_2}
	\frac{\delta S_{MCS1}}{\delta\Phi}=0\implies\frac{\delta S}{\delta\Phi}=-\int d^7z\left[\left(G-g^2k\right)\frac{\delta k}{\delta\Phi}+\left(\Pi_{\frac{1}{2}}V\right)\frac{\delta J}{\delta\Phi}\right].
\end{equation}
The inhomogeneous Klein-Gordon equation can also be derived from (\ref{G_eq}) by applying the operator $\bar D^\alpha D_\alpha-m$ to both sides of (\ref{G_eq}) and using the identity $\Pi_{\frac{1}{2}}G=G$. This yields the result
\begin{equation}
	\left(\Box-m^2\right)G=g^2\left(-mJ+\bar D^\alpha D_\alpha J-m\bar D^\alpha D_\alpha k+\Box\Pi_{\frac{1}{2}} k\right)
\end{equation}
or
\begin{equation}
	\label{G_solved}
	G=g^2\hat{\mathcal{O}}\left(-mJ+\bar D^\alpha D_\alpha J-m\bar D^\alpha D_\alpha k+\Box\Pi_{\frac{1}{2}} k\right).
\end{equation}
This leads us to
\begin{equation}
	\label{Pi_V_solved}
	\Pi_{\frac{1}{2}}V=g^2\hat{\mathcal{O}}\left(\bar D^\alpha D_\alpha k-m\Pi_{\frac{1}{2}}k+J-m\frac{\bar D^\alpha D_\alpha}{\Box} J\right). 
\end{equation}
Finally, substituting (\ref{G_solved}) and (\ref{Pi_V_solved}) into (\ref{matter_2}), we obtain after some algebraic work
\begin{equation}
	\label{final_matter_eq2}
	\begin{split}
		\frac{\delta S}{\delta\Phi}=&g^2\int d^7z\bigg[\hat{\mathcal{O}}\left(mJ-\bar D^\alpha D_\alpha J-m^2k+m\bar D^\alpha D_\alpha k+\{D^2,\bar D^2\}k\right)\frac{\delta k}{\delta\Phi}\\
		&+\left(-\bar D^\alpha D_\alpha k+m\Pi_{\frac{1}{2}}k-J+\frac{m}{\Box}\bar D^\alpha D_\alpha J\right)\frac{\delta J}{\delta\Phi}\bigg].
	\end{split}
\end{equation}
It is evident that this field equation is identical to the one in (\ref{final_matter_eq1}), which implies that the matter sectors of $S_{SD1}$ and $S_{MCS1}$ give rise to the same dynamics. This completes the demonstration of the dual equivalence between the scalar self-dual and MCS models coupled to matter at the level of field equations.

Let us now proceed to define our second master action $S_{M2}$. It is defined as
\begin{equation}
	\begin{split}
		\label{master_2}
		S_{M2}=&\frac{1}{2g^2}\left[\int d^5z\left(\pi^\alpha W_\alpha-\frac{1}{2}\pi^\alpha\pi_\alpha\right)+\int d^5\bar z\left(\bar \pi^\alpha\bar W_\alpha-\frac{1}{2}\bar\pi^\alpha\bar\pi_\alpha\right)-\int d^7z mVG\right]\\
		&+\int d^5z\pi^\alpha k_\alpha+\int d^5\bar z\bar\pi^\alpha \bar k_\alpha+\int d^7zVJ+S[\bar\Phi,\Phi],
	\end{split}
\end{equation}
where the superfield strengths $W^\alpha$ and $\bar W^\alpha$ are expressed in terms of $G$ as \cite{BPS}
\begin{equation}
	W_\alpha=-i\bar D_\alpha G=i\bar D^2D_\alpha V \ \ \text{and} \ \ \bar W_\alpha=iD_\alpha G=-iD^2\bar D_\alpha V,
\end{equation}
and they satisfy the condition
\begin{equation}
	D^\alpha W_\alpha+\bar	D^\alpha\bar W_\alpha=0.
\end{equation}
Although $S_{M2}$ shares certain similarities with $S_{M1}$, it is important to note that, in addition to $V$, $\Phi$, and $\bar \Phi$, the second master action is formulated in terms of a spinor chiral superfield $\pi^\alpha$, which we will refer to as the spinor self-dual superfield. This superfield interacts with matter through a minimal coupling involving a current $k_\alpha$. The master action $S_{M2}$ is invariant under local $U(1)$ transformations, if $J$ satisfies the linearity condition $J=\Pi_\frac{1}{2}J$ and only the gauge superfield $V$ undergoes a non-trivial transformation: $\delta V=i\left(\bar \Lambda-\Lambda\right)$.

In order to derive the MCS action from (\ref{master_2}), it is first necessary to consider the variation of (\ref{master_2}) with respect to $\pi^\alpha$ and $\bar\pi^\alpha$. This yields the field equations
\begin{equation}
	\label{duality_condition}
	\pi_\alpha=W_\alpha+2g^2 k_\alpha \ \ ; \ \ \bar \pi_\alpha=\bar W_\alpha+2g^2\bar k_\alpha.
\end{equation}
Similar to (\ref{identification1}), if we set the currents $k_\alpha=0$ and $\bar k_\alpha=0$, we obtain the self-duality conditions $\pi_\alpha=W_\alpha$ and $\bar \pi_\alpha=\bar W_\alpha$. Once the expressions (\ref{duality_condition}) have been inserted back into (\ref{master_2}), the result is
\begin{equation}
	\begin{split}
		\label{MCS2}
		S_{MCS2}=&-\frac{1}{2g^2}\int d^7z\left(G^2+mVG\right)+\int d^5z \left(W^\alpha+g^2k^\alpha\right)k_\alpha\\
		&+\int d^5\bar z \left(\bar W^\alpha+g^2\bar k^\alpha\right)\bar k_\alpha+\int d^7zVJ+S[\bar\Phi,\Phi].
	\end{split}
\end{equation} 
It is worth noting that the minimal couplings $\pi^\alpha k_\alpha$ and $\bar\pi^\alpha\bar k_\alpha$ in (\ref{master_2}) were replaced by non-minimal couplings in $S_{MCS2}$. Furthermore, $S_{MCS2}$ is gauge invariant and very similar to (\ref{MCS1}), with the only difference being that the magnetic-like coupling employs the spinor superfield strength $W^\alpha$ (and $\bar W^\alpha$), rather than the scalar one $G$. We note that, in analogy with above discussions, this model is an extension of the free $\mathcal{N} = 2$ MCS action (\ref{MCS20}), with currents added. Finally, it is important to point out that the two $\mathcal{N}=2$ MCS models, (\ref{MCS1}) and (\ref{MCS2}), are distinct due to their slightly different interactions. Nevertheless, if we assume that $k$ is a linear superfield ($D^2k=0$ and $\bar D^2k=0$), it is not difficult to demonstrate that $S_{MCS1}=S_{MCS2}$ when the identifications $k_\alpha=\displaystyle\frac{i}{2}\bar D_\alpha k$ and $\bar k_\alpha=-\displaystyle\frac{i}{2}D_\alpha k$ are implemented. 

On the other hand, the variation of (\ref{master_2}) with respect to $V$ leads to the field equation
\begin{equation}
	\label{G2}
	G=-\frac{i}{2m}\left(D^\alpha\pi_\alpha-\bar D^\alpha\bar\pi_\alpha\right)+\frac{g^2}{m}J.
\end{equation}
From this, we can show that
\begin{align}
	\label{PIV2}
	\Pi_{\frac{1}{2}}V&=\frac{1}{2m\Box}\left(\partial_\alpha^{\phantom{\alpha}\beta}D^\alpha\pi_\beta-\partial_\alpha^{\phantom{\alpha}\beta}\bar D^\alpha\bar\pi_\beta\right)+\frac{g^2}{m}\frac{\bar D^\alpha D_\alpha}{\Box}J;\\
	W_\alpha&=-\frac{1}{2m}\left(i\partial_\alpha^{\phantom{\alpha}\beta}\pi_\beta+\bar D^2\bar\pi_\alpha\right)-i\frac{g^2}{m}\bar D_\alpha J;\\
	\label{antichiral}
	\bar W_\alpha&=-\frac{1}{2m}\left(i\partial_\alpha^{\phantom{\alpha}\beta}\bar\pi_\beta+D^2\pi_\alpha\right)+i\frac{g^2}{m} D_\alpha J.
\end{align}
Again, due to the identities (\ref{identity}), we can substitute (\ref{G2}-\ref{antichiral}) into (\ref{master_2}) to obtain (after a series of algebraic manipulations) the $\mathcal{N} = 2$ spinor self-dual action coupled to matter:
\begin{equation}
	\begin{split}
		\label{SD2}
		S_{SD2}=&-\frac{1}{4mg^2}\bigg\{\int d^7z\bar \pi^\alpha\pi_\alpha+\int d^5z\left[\frac{1}{2}\pi^\alpha i{\partial_\alpha}^\beta\pi_\beta+m\pi^\alpha\pi_\alpha\right]\\
		&+\int d^5\bar{z}\left[\frac{1}{2}\bar \pi^\alpha i{\partial_\alpha}^\beta\bar \pi_\beta+m\bar\pi^\alpha\bar\pi_\alpha\right]\bigg\}+\int d^5z\pi^\alpha\left( k_\alpha-\frac{i}{2m}\bar{D}_\alpha J\right)\\
		&+\int d^5\bar z\bar \pi^\alpha\left( \bar k_\alpha+\frac{i}{2m} D_\alpha J\right)+\int d^7z\frac{g^2}{2m} J\frac{\bar D^\alpha D_\alpha}{\Box}J+S[\bar \Phi,\Phi].
	\end{split}
\end{equation}
In this instance, the minimal coupling $VJ$ in (\ref{master_2}) was replaced by non-minimal couplings in $S_{SD2}$. Furthermore, the non-local Thirring-like interaction is identical to that given in (\ref{SD1}). Since $S_{MCS2}$ and $S_{SD2}$ originate from the same master action $S_{M2}$ as defined in (\ref{master_2}), it can be concluded that $S_{MCS2}$ and $S_{SD2}$ are mutually dual. Once more, in analogy with above discussions, this model is an extension of the free $\mathcal{N} = 2$ SD action (\ref{SD20}), with currents added.

The analysis of the field equations derived from models $S_{MCS2}$ and $S_{SD2}$ will now be conducted in order to reveal the dualization between $S_{MCS2}$ and $S_{SD2}$. The first step is to vary (\ref{MCS2}) with respect to $V$, which leads to
\begin{equation}
	\label{G_2}
	\bar D^\alpha D_\alpha G+mG=ig^2\bar D^\alpha \bar k_\alpha-ig^2D^\alpha k_\alpha+g^2J.
\end{equation}
By applying $-i\bar D_\alpha$ to both sides of (\ref{G_2}), an equation for $W_\alpha$ is obtained. Similarly, by applying $iD_\alpha$ to both sides of (\ref{G_2}), an equation for $\bar W_\alpha$ is derived. These equations are
\begin{align}
	\label{W1}
		i{\partial_\alpha}^\beta W_\beta+mW_\alpha&=-g^2\left(i{\partial_\alpha}^\beta k_\beta+\bar D^2 \bar k_\alpha+i\bar D_\alpha J\right);\\
		\label{W2}
		i{\partial_\alpha}^\beta \bar W_\beta+m\bar W_\alpha&=-g^2\left(i{\partial_\alpha}^\beta \bar k_\beta+D^2 k_\alpha-iD_\alpha J\right).
\end{align}
Consequently, the dynamics of the spinor superfield strengths are determined by inhomoge- neous massive Weyl equations.

The second step is to vary (\ref{SD2}) with respect to $\pi^\alpha$ and $\bar\pi^\alpha$. This leads to
\begin{align}
	\label{eq1}
	\bar D^2\bar \pi_\alpha+ i{\partial_\alpha}^\beta \pi_\beta+2m\pi_\alpha&=4mg^2k_\alpha-2ig^2\bar{D}_\alpha J;\\
	\label{eq2}
	D^2 \pi_\alpha+ i{\partial_\alpha}^\beta\bar \pi_\beta+2m\bar \pi_\alpha&=4mg^2\bar k_\alpha+2ig^2 D_\alpha J.
\end{align}
It is convenient to demonstrate that the superfields $\pi^\alpha$ and $\bar\pi^\alpha$ satisfy inhomogeneous massive Weyl equations. To achieve this, we begin by contracting equation (\ref{eq1}) with $D^\alpha$ and (\ref{eq2}) with $\bar D^\alpha$, and then we use the identities $D^\alpha\bar D^2\bar \pi_\alpha=-i{\partial_\beta}^\alpha\bar D^\beta\bar \pi_\alpha$ and $\bar D^\alpha D^2\pi_\alpha=-i{\partial_\beta}^\alpha D^\beta\pi_\alpha$. The resulting equations are then added to yield a constraint
\begin{equation}
	D^\alpha\pi_\alpha+\bar D^\alpha\bar \pi_\alpha=2g^2\left(D^\alpha k_\alpha+\bar D^\alpha\bar k_\alpha\right).
\end{equation} 
By substituting $\bar D^2\bar \pi_\alpha=-\bar D_\alpha\bar D^\beta \bar \pi_\beta$ and $ D^2 \pi_\alpha=- D_\alpha D^\beta  \pi_\beta$ into equations (\ref{eq1}) and (\ref{eq2}), and employing the constraint that has just been derived, we obtain 
\begin{align}
	\label{pi1}
	i{\partial_\alpha}^\beta \pi_\beta+m\pi_\alpha&=g^2\left(i{\partial_\alpha}^\beta k_\beta+2mk_\alpha-\bar D^2\bar k_\alpha-i\bar{D}_\alpha J\right);\\
	\label{pi2}
	i{\partial_\alpha}^\beta \bar \pi_\beta+m\bar \pi_\alpha&=g^2\left(i{\partial_\alpha}^\beta \bar k_\beta+2m\bar k_\alpha- D^2 k_\alpha+iD_\alpha J\right).
\end{align}
A direct examination of the superfield equations (\ref{W1}-\ref{W2}) and (\ref{pi1}-\ref{pi2}) reveals that they are identical when the identifications (\ref{duality_condition}) are made.

In the following, it is our aim to derive the field equations for the matter superfields from $S_{SD2}$ and $S_{MCS2}$. We begin with $S_{SD2}$. If we now vary $(\ref{SD2})$ with respect to $\Phi$, we find that $\Phi$ obeys the field equation
\begin{equation}
	\label{SD2Phi}
	\begin{split}
		\frac{\delta S_{SD2}}{\delta\Phi}=0\Rightarrow\frac{\delta S}{\delta\Phi}=&-\int d^5z\pi^\alpha\left(\frac{\delta k_\alpha}{\delta\Phi}-\frac{i}{2m}\bar D_\alpha\frac{\delta J}{\delta\Phi}\right)\\
		&-\int d^5\bar z\bar\pi^\alpha\left(\frac{\delta k_\alpha}{\delta\Phi}+\frac{i}{2m} D_\alpha\frac{\delta J}{\delta\Phi}\right)-\int d^7z\frac{g^2}{m}\frac{\bar D^\alpha D_\alpha}{\Box}J\frac{\delta J}{\delta\Phi}.
	\end{split}
\end{equation}
The application of the linear operator $i{\partial_\gamma}^\alpha-{\delta_\gamma}^\alpha m$ to both sides of (\ref{pi1}) and (\ref{pi2}) leads to inhomogeneous Klein-Gordon equations whose solutions are given by
\begin{align}
	\pi_\gamma&=g^2\hat{\mathcal{O}}_\gamma^{\phantom\alpha\alpha}\left(i{\partial_\alpha}^\beta k_\beta+2mk_\alpha-\bar D^2\bar k_\alpha-i\bar{D}_\alpha J\right);\\
	\bar{ \pi}_\gamma&=g^2\hat{\mathcal{O}}_\gamma^{\phantom\alpha\alpha}\left(i{\partial_\alpha}^\beta \bar k_\beta+2m\bar k_\alpha- D^2 k_\alpha+iD_\alpha J\right),
\end{align}
where
\begin{equation}
	\hat{\mathcal{O}}_\gamma^{\phantom\alpha\alpha}:=\frac{i{\partial_\gamma}^\alpha-{\delta_\gamma}^\alpha m}{\Box-m^2}.
\end{equation}
Substituting these solutions into (\ref{SD2Phi}), we obtain
\begin{align}
	\frac{\delta S}{\delta\Phi}=&\int d^5zg^2\hat{\mathcal{O}}^{\alpha\gamma}\left(\bar D^2\bar k_\gamma+i\bar{D}_\gamma J-i{\partial_\gamma}^\beta k_\beta-2mk_\gamma\right)\left(\frac{\delta k_\alpha}{\delta\Phi}-\frac{i}{2m}\bar D_\alpha\frac{\delta J}{\delta\Phi}\right)\nonumber\\
	&+\int d^5\bar zg^2\hat{\mathcal{O}}^{\alpha\gamma}\left(D^2 k_\gamma-iD_\gamma J-i{\partial_\gamma}^\beta \bar k_\beta-2m\bar k_\gamma\right)\left(\frac{\delta k_\alpha}{\delta\Phi}+\frac{i}{2m} D_\alpha\frac{\delta J}{\delta\Phi}\right)\nonumber\\
	&-\int d^7z\frac{g^2}{m}\frac{\bar D^\alpha D_\alpha}{\Box}J\frac{\delta J}{\delta\Phi},
\end{align}
where $\hat{\mathcal{O}}^{\alpha\gamma}=\hat{\mathcal{O}}\left(i\partial^{\alpha\gamma}-mC^{\alpha\gamma}\right)$.

It is convenient to rewrite the integrals that involve the distribution $\frac{\delta J}{\delta\Phi}$ and are over the chiral and anti-chiral subspaces as an integral over the full superspace. Following a series of algebraic manipulations, it can be demonstrated that
\begin{equation}
	\label{finaleqphi}
	\begin{split}
		\frac{\delta S}{\delta\Phi}=&\int d^5zg^2\hat{\mathcal{O}}^{\alpha\gamma}\left(\bar D^2\bar k_\gamma+i\bar{D}_\gamma J-i{\partial_\gamma}^\beta k_\beta-2mk_\gamma\right)\frac{\delta k_\alpha}{\delta\Phi}\\
		&+\int d^5\bar zg^2\hat{\mathcal{O}}^{\alpha\gamma}\left(D^2 k_\gamma-iD_\gamma J-i{\partial_\gamma}^\beta \bar k_\beta-2m\bar k_\gamma\right)\frac{\delta\bar k_\alpha}{\delta\Phi}\\
		&-\int d^7zg^2\hat{\mathcal{O}}\big(i\bar D^\alpha\bar k_\alpha-iD^\alpha k_\alpha+\frac{m}{\Box}\partial^{\alpha\beta} D_\beta k_\alpha-\frac{m}{\Box}\partial^{\alpha\beta}\bar D_\beta \bar k_\alpha\\
		&-\frac{m}{\Box}\bar D^\gamma{D}_\gamma J+J\big)\frac{\delta J}{\delta\Phi}.
	\end{split}
\end{equation}
Now, let us vary $S_{MCS2}$ with respect to $\Phi$ to derive the following superfield equation
\begin{align}
	\label{SD2Phi_2}
	\frac{\delta S_{MCS2}}{\delta\Phi}=0\Rightarrow\frac{\delta S}{\delta\Phi}=&-\int d^5z\left(W^\alpha+2g^2k^\alpha\right)\frac{\delta k_\alpha}{\delta\Phi}-\int d^5\bar z\left(\bar W^\alpha+2g^2\bar k^\alpha\right)\frac{\delta\bar k_\alpha}{\delta\Phi}\nonumber\\
	&-\int d^7z\left(\Pi_{\frac{1}{2}}V\right)\frac{\delta J}{\delta\Phi}.
\end{align}
The application of $\bar D^\beta D_\beta- m$ to both sides of (\ref{G_2}) enable us to express $G$ exclusively in terms of current terms, as follows:
\begin{equation}
	G=g^2\hat{\mathcal{O}}\left(\partial_\beta^{\phantom{\beta}\alpha}D^\beta k_\alpha-\partial_\beta^{\phantom{\beta}\alpha}\bar D^\beta \bar k_\alpha+\bar D^\beta D_\beta J- mi\bar D^\alpha \bar k_\alpha+miD^\alpha k_\alpha-mJ\right).
\end{equation}
This result implies
\begin{align}
	\label{W11}
	W^\alpha+2g^2k^\alpha&=-g^2\hat{\mathcal{O}}^{\alpha\gamma}\left(\bar D^2\bar k_\gamma+i\bar{D}_\gamma J-i{\partial_\gamma}^\beta k_\beta-2mk_\gamma\right);\\
	\label{W22}
	\bar W^\alpha+2g^2\bar k^\alpha&=-g^2\hat{\mathcal{O}}^{\alpha\gamma}\left(D^2 k_\gamma-iD_\gamma J-i{\partial_\gamma}^\beta \bar k_\beta-2m\bar k_\gamma\right);\\
	\label{PiV}
	\Pi_{\frac{1}{2}}V&=g^2\hat{\mathcal{O}}\Big(i\bar D^\alpha \bar k_\alpha-iD^\alpha k_\alpha+J+\frac{m}{\Box}\partial_\beta^{\phantom{\beta}\alpha}\bar D^\beta \bar k_\alpha-\frac{m}{\Box}\partial_\beta^{\phantom{\beta}\alpha} D^\beta k_\alpha\nonumber\\
	&-m\frac{\bar D^\beta D_\beta}{\Box} J\Big).
\end{align}
Finally, by substituting (\ref{W11}-\ref{PiV}) into (\ref{SD2Phi_2}) we obtain exactly the field equation (\ref{finaleqphi}). This means that the matter sectors of $S_{MCS2}$ in (\ref{MCS2}) and $S_{SD2}$ in (\ref{SD2}) are equivalent. Consequently, we have thoroughly established the dual equivalence between the  MCS and spinor self-dual models in the presence of dynamical matter chiral superfields at the level of field equations.

\section{Equivalence at the level of generating functionals}

This study has thus far focused on duality at the level of field equations, that is, at the classical level. What follows will address duality at the level of generating functionals, that is, at the quantum level. The generating functional for the Green functions of the master theory given in (\ref{master_1}) is defined as:
\begin{equation}
	\label{Z1master}
	\begin{split}
		\mathcal{Z}_1=&N_1\int \mathcal{D}V\mathcal{D}\sigma\exp\bigg\{\frac{1}{g^2}\int d^7z \left(-\sigma G+\frac{1}{2}\sigma^2-\frac{m}{2}VG\right)+\int d^7z\left(\sigma k+VJ\right)\\
		&+S[\bar \Phi,\Phi]\bigg\},
	\end{split}
\end{equation}
where the normalization constant $N_1$ is introduced to absorb field-independent factors.

Let us consider changing the integration variables $V\rightarrow V-\frac{1}{m}\sigma+\frac{g^2}{m}\frac{\bar D^\alpha D_\alpha}{\Box}J$ and $\sigma\rightarrow\sigma$. First, the Jacobian determinant of such transformation is equal to one. Second, the superfield $V$ is completely
decoupled. As a result, the contribution of the integral over $V$ to $\mathcal{Z}_1$ is only an overall constant, which can be absorbed into $N_1$. Therefore, Eq. (\ref{Z1master}) implies
\begin{equation}
	\label{Z1dual}
	\begin{split}
		\mathcal{Z}_1=&N_1\int\mathcal{D}\sigma\exp\left(S_{SD1}\right),
	\end{split}
\end{equation}
where $S_{SD1}$ is the scalar self-dual action obtained in Eq. (\ref{SD1}).
  
Alternatively, if the change of variables $V\rightarrow V$ and $\sigma\rightarrow\sigma+G-g^2k$ is considered, the corresponding Jacobian is equal to one. However, the superfield $\sigma$ is now completely decoupled. Therefore, the integral over $\sigma$ can be incorporated into $N_1$, as it is an overall constant. It can thus be concluded from (\ref{Z1master}) that  
\begin{equation}
	\label{Z1MCS}
	\begin{split}
		\mathcal{Z}_1=&N_1\int \mathcal{D}V\exp\left(S_{MCS1}\right),
	\end{split}
\end{equation}
where $S_{MCS1}$ is the MCS model found in (\ref{MCS1}).

We have demonstrated that, with the appropriate change of variables, the generating functional (\ref{Z1master}) can be expressed as a generating functional for the scalar self-dual model (\ref{Z1dual}) or a generating functional for the MCS theory (\ref{Z1MCS}). Therefore, the dual equivalence of the scalar self-dual model (\ref{SD1}) and the MCS theory (\ref{MCS1}) holds at the quantum level. This implies that the two models represent the same physics, but that the physical description is given using different variables.  

We will conclude our analysis by performing the functional integral over $\sigma$ in Eq. (\ref{Z1dual}) and the integral over $V$ in (\ref{Z1MCS}). We will begin with Eq. (\ref{Z1dual}): 
\begin{equation}
	\begin{split}
		N_1\int\mathcal{D}\sigma\exp\left(S_{SD1}\right)&=N_1\exp\bigg\{\int d^7z\frac{g^2}{2m}J\frac{\bar D^\alpha D_\alpha}{\Box}J+S[\bar \Phi,\Phi]\bigg\}\int\mathcal{D}\sigma\\
		&\times\exp\bigg\{\frac{1}{2g^2}\int d^7z \sigma\left(\frac{1}{m}\bar D^\alpha D_\alpha+1\right)\sigma+\int d^7z\sigma\left( k-\frac{1}{m}J\right)\bigg\}.
	\end{split}
\end{equation}
This is a standard Gaussian integral that can be evaluated with the help of the identity
\begin{equation}
	\left[\frac{1}{g^2}\left(\frac{1}{m}\bar D^\alpha D_\alpha+1\right)\right]^{-1}=g^2\left(m\mathcal{O}\bar D^\alpha D_\alpha-m^2\mathcal{O}\Pi_{\frac{1}{2}}+\Pi_0\right).
\end{equation}
Therefore, after some algebraic work, we find the result
\begin{equation}
	\label{final_SD1}
	\begin{split}
		N_1\int\mathcal{D}\sigma\exp\left(S_{SD1}\right)=&N_1^\sigma\exp\left(S_{eff1}[\bar{\Phi},\Phi]\right),
	\end{split}
\end{equation}
where field-independent factors have been absorbed by the constant $N_1^\sigma$ and
\begin{equation}
	\label{effective_1}
	\begin{split}
		S_{eff1}:=& \ S[\bar \Phi,\Phi]-\int d^7z\frac{g^2}{2}\bigg[k\left(m\mathcal{O}\bar D^\alpha D_\alpha-m^2\mathcal{O}\Pi_{\frac{1}{2}}+\Pi_0\right)k\\
		&+2k\left(-\mathcal{O}\bar D^\alpha D_\alpha+m\mathcal{O}\right)J+J\left(\frac{m}{\Box}\mathcal{O}\bar D^\alpha D_\alpha-\mathcal{O}\right)J\bigg].
	\end{split}
\end{equation}
It is noteworthy that the variation of the effective matter action $S_{eff1}$ with respect to $\Phi$ yields the same field equations (\ref{final_matter_eq1}) and (\ref{final_matter_eq2}) that were obtained in the classical case.

We now turn to the calculation of the integral over $V$ in Eq. (\ref{Z1MCS}). Due to the gauge invariance of $S_{SMC1}$, the integral in (\ref{Z1MCS}) gives an infinite contribution to $\mathcal{Z}_1$. This issue can be overcome by adding to $S_{SMC1}$ some gauge-fixing term. We choose the standard one
\begin{equation}
	\label{GF}
	S_{GF}=-\frac{1}{g^2\xi}\int d^7z \left(\bar D^2 V\right)D^2 V,
\end{equation}
where $\xi$ is a gauge-fixing parameter.

Thus, the generating functional for $S_{SMC1}+S_{GF}$ is given by
\begin{align}
	\begin{split}
		N_1&\int \mathcal{D}V\exp\left(S_{MCS1}+S_{GF}\right)=N_1\exp\bigg\{-\int d^7z\frac{g^2}{2}k^2+S[\bar \Phi,\Phi]\bigg\}\int \mathcal{D}V\\
		\times&\exp\bigg\{-\frac{1}{2g^2}\int d^7z V\bigg(\Box\Pi_{\frac{1}{2}}+m\bar D^\alpha D_\alpha+\frac{1}{\xi}\Pi_0\bigg)V+\int d^7zV\left(J+\bar D^\alpha D_\alpha k\right)\bigg\}.
	\end{split}
\end{align}
This integral can be evaluated with the help of the following identity:
\begin{equation}
	\label{propagator}
	\left[-\frac{1}{g^2}\left(\Box\Pi_{\frac{1}{2}}+m\bar D^\alpha D_\alpha+\frac{1}{\xi}\Pi_0\right)\right]^{-1}=-g^2\left(\mathcal{O}\Pi_{\frac{1}{2}}-\frac{m}{\Box}\mathcal{O}\bar D^\alpha D_\alpha+\xi\Pi_0\right).
\end{equation}
Therefore,
\begin{equation}
		\begin{split}
		N_1&\int \mathcal{D}V\exp\left(S_{MCS1}+S_{GF}\right)=N_1^V\exp\bigg\{-\int d^7z\frac{g^2}{2}k^2+S[\bar \Phi,\Phi]\bigg\}\\
		&\times\exp\bigg\{\frac{g^2}{2}\int d^7z \left(J+\bar D^\beta D_\beta k\right)\left(\mathcal{O}\Pi_{\frac{1}{2}}-\frac{m}{\Box}\mathcal{O}\bar D^\alpha D_\alpha+\xi\Pi_0\right)\left(J+\bar D^\gamma D_\gamma k\right)\bigg\},
	\end{split}
\end{equation}
where $N_1^V$ is a normalization constant.

Given that $\Pi_0\left(J+\bar D^\gamma D_\gamma k\right)=0$, it follows that the result is independent of the parameter $\xi$. Consequently, 
\begin{equation}
	\label{final_MCS1}
	N_1\int \mathcal{D}V\exp\left(S_{MCS1}+S_{GF}\right)=N_1^V\exp\left(S_{eff1}[\Bar\Phi,\Phi]\right),
\end{equation}
where $S_{eff1}[\Bar\Phi,\Phi]$ is defined in (\ref{effective_1}).

It can be observed in (\ref{final_SD1}) and (\ref{final_MCS1}) that the generating functionals (\ref{Z1dual}) and (\ref{Z1MCS}) yield the same result, with the exception of field-independent factor. This serves to validate the assertion that the generating functional for the scalar self-dual model and the generating functional for the MCS model are, in fact, the same generating functional, but expressed with different variables.

Let us now proceed to define the generating functional for the master action $S_{M2}$, as defined in (\ref{master_2}). The functional is defined as follows:
\begin{equation}
	\label{Z2master}
	\begin{split}
		\mathcal{Z}_2=&N_2\int \mathcal{D}V\mathcal{D}\pi^\alpha\mathcal{D}\bar\pi^\alpha\exp\bigg\{\frac{1}{2g^2}\bigg[\int d^5z\left(\pi^\alpha W_\alpha-\frac{1}{2}\pi^\alpha\pi_\alpha\right)+\int d^5\bar z\bigg(\bar\pi^\alpha \bar W_\alpha\\
		&-\frac{1}{2}\bar\pi^\alpha\bar\pi_\alpha\bigg)-\int d^7zmVG\bigg]+\int d^5z\pi^\alpha k_\alpha+\int d^5\bar z\bar\pi^\alpha\bar k_\alpha+\int d^7zVJ+S[\bar \Phi,\Phi]\bigg\}.
	\end{split}
\end{equation}
In order to completely decouple the superfield $V$, we begin by defining the following change of variables: $V\rightarrow V+\frac{1}{2m\Box}\left(\partial^{\phantom{\alpha}\beta}_\alpha D^\alpha\pi_\beta-\partial^{\phantom{\alpha}\beta}_\alpha \bar D^\alpha\bar\pi_\beta\right)+\frac{g^2}{m}\frac{\bar D^\alpha D_\alpha}{\Box}J$, $\pi^\alpha\rightarrow\pi^\alpha$, and $\bar \pi^\alpha\rightarrow\bar\pi^\alpha$. After performing a series of algebraic manipulations and incorporating the integral over $V$ into $N_2$, the generating functional (\ref{Z2master}) can be rewritten as follows:
\begin{equation}
	\label{Z2SD2}
	\mathcal{Z}_2=N_2\int\mathcal{D}\pi^\alpha\mathcal{D}\bar\pi^\alpha
	\exp\left(S_{SD2}\right),
\end{equation}
where $S_{SD2}$ is the spinor self-dual action found in Eq. (\ref{SD2}).

On the other hand, if the change of variables $V\rightarrow V$, $\pi^\alpha\rightarrow\pi^\alpha+W^\alpha+2g^2k^\alpha$, and $\bar\pi^\alpha\rightarrow\bar\pi^\alpha+\bar W^\alpha+2g^2\bar k^\alpha$ is considered, the complete decoupling of the superfields $\pi^\alpha$ and $\bar \pi^\alpha$ is achieved.  Consequently, we can rewrite (\ref{Z2master}) as follows:
\begin{equation}
	\label{Z2MCS2}
	\mathcal{Z}_2=N_2\int \mathcal{D}V\exp\left(S_{MCS2}\right),
\end{equation}
where $S_{MCS2}$ is the MCS model obtained in Eq. (\ref{MCS2}).

The demonstration that the generating functional for the spinor self-dual model (\ref{Z2SD2}) and the generating functional for the MCS model (\ref{Z2MCS2}) are, in fact, identical to (\ref{Z2master}), but expressed using different variables, substantiates the dual equivalence between the spinor self-dual model (\ref{SD2}) and the MCS theory (\ref{MCS2}) at the quantum level.

The analysis will be concluded by performing the functional integral over $\pi^\alpha$ and $\bar\pi^\alpha$ in Eq. (\ref{Z2SD2}) and the integral over $V$ in (\ref{Z2MCS2}). To begin, it is convenient to introduce the projection operators for (anti-)chiral spinor superfields: 
\begin{equation}
	\label{projection}
	P_\pm:=\frac{1}{2}(1\pm \bf K),
\end{equation}
where ${\bf K}$ is the rest-frame conjugation operator \cite{GGRS,SG}, which by definition acts on $\pi^\alpha$ and $\bar\pi^\alpha$ as follows:
\begin{equation}
	{\bf K}\pi_\alpha:=-\frac{i\partial^{\phantom{\alpha}\beta}_\alpha}{\Box}\bar D^2\bar\pi_\beta \ \ \ ; \ \ \ {\bf K}\bar\pi_\alpha:=-\frac{i\partial^{\phantom{\alpha}\beta}_\alpha}{\Box} D^2\pi_\beta.
\end{equation}
With these definitions in hand, we can write (\ref{Z2SD2}) as
\begin{equation}
	\label{interm}
	\begin{split}
		N_2&\int\mathcal{D}\pi^\alpha\mathcal{D}\bar\pi^\alpha
		\exp\left(S_{SD2}\right)=N_2\exp\bigg\{\int d^7z\frac{g^2}{2m}J\frac{\bar D^\alpha D_\alpha}{\Box}J+S[\bar \Phi,\Phi]\bigg\}\\
		&\times\int\mathcal{D}\pi^\alpha\exp\bigg\{-\frac{1}{4mg^2}\int d^7z\pi^\alpha\left[m\delta^{\phantom{\alpha}\beta}_\alpha P_++\left(i\partial^{\phantom{\alpha}\beta}_\alpha +m\delta^{\phantom{\alpha}\beta}_\alpha\right) P_-\right]\pi_\beta\\
		&+\int d^5z\pi^\alpha\left(k_\alpha-\frac{i}{2m}\bar D_\alpha J\right)\bigg\}\\
		&\times\int\mathcal{D}\bar\pi^\alpha\exp\bigg\{-\frac{1}{4mg^2}\int d^7z\bar\pi^\alpha\left[m\delta^{\phantom{\alpha}\beta}_\alpha P_++\left(i\partial^{\phantom{\alpha}\beta}_\alpha +m\delta^{\phantom{\alpha}\beta}_\alpha\right) P_-\right]\bar\pi_\beta\\
		&+\int d^5\bar z\bar\pi^\alpha\left(\bar k_\alpha+\frac{i}{2m} D_\alpha J\right)\bigg\}.
	\end{split}
\end{equation}
Given their properties of completeness, idempotence, and orthogonality, the projection operators permit the inversion of the differential operator in Eq. (\ref{interm}) with relative ease. The inverse operator is
\begin{equation}
	\left\{-\frac{1}{2mg^2}\left[m\delta^{\phantom{\alpha}\beta}_\alpha P_++\left(i\partial^{\phantom{\alpha}\beta}_\alpha +m\delta^{\phantom{\alpha}\beta}_\alpha\right) P_-\right]\right\}^{-1}=-2mg^2\left[\frac{1}{m}\delta^{\phantom{\alpha}\beta}_\alpha P_++\frac{i\partial^{\phantom{\alpha}\beta}_\alpha -m\delta^{\phantom{\alpha}\beta}_\alpha}{\Box-m^2} P_-\right]
\end{equation}
It follows that the Gaussian integrals over $\pi^\alpha$ and $\bar \pi^\alpha$ can now be evaluated to yield
\begin{equation}
	\begin{split}
		N_2&\int\mathcal{D}\pi^\alpha\mathcal{D}\bar\pi^\alpha
		\exp\left(S_{SD2}\right)=N_2^\pi\exp\bigg\{\int d^7z\frac{g^2}{2m}J\frac{\bar D^\alpha D_\alpha}{\Box}J+S[\bar \Phi,\Phi]\bigg\}\\
		&\times\exp\bigg\{mg^2\int d^5z\left(k^\alpha-\frac{i}{2m}\bar D^\alpha J\right)\left[\frac{1}{m}\delta^{\phantom{\alpha}\beta}_\alpha P_++\frac{i\partial^{\phantom{\alpha}\beta}_\alpha -m\delta^{\phantom{\alpha}\beta}_\alpha}{\Box-m^2} P_-\right]\left(k_\beta-\frac{i}{2m}\bar D_\beta J\right)\bigg\}\\
		&\times\exp\bigg\{mg^2\int d^5\bar z\left(\bar k^\alpha+\frac{i}{2m} D^\alpha J\right)\left[\frac{1}{m}\delta^{\phantom{\alpha}\beta}_\alpha P_++\frac{i\partial^{\phantom{\alpha}\beta}_\alpha -m\delta^{\phantom{\alpha}\beta}_\alpha}{\Box-m^2} P_-\right]\left(\bar k_\beta+\frac{i}{2m} D_\beta J\right)\bigg\},
	\end{split}
\end{equation}
where field-independent factors have been incorporated into the constant $N_2^\pi$.

Since $\bar D_\beta J$ is a chiral spinor superfield and $D_\beta J$ is a anti-chiral one, we have
\begin{equation}	
	P_\pm\left(-\frac{i}{2m}\bar D_\beta J\right)=
	\begin{cases}
		\displaystyle0\\
		\displaystyle-\frac{i}{2m}\bar D_\beta J
	\end{cases} \ \ \ ; \ \ \ 	P_\pm\left(\frac{i}{2m}D_\beta J\right)=
	\begin{cases}
	\displaystyle0\\
	\displaystyle\frac{i}{2m}D_\beta J
	\end{cases}.
\end{equation}
Therefore,
\begin{equation}
	\begin{split}
		N_2&\int\mathcal{D}\pi^\alpha\mathcal{D}\bar\pi^\alpha
		\exp\left(S_{SD2}\right)=N_2^\pi\exp\bigg\{\int d^7z\frac{g^2}{2m}J\frac{\bar D^\alpha D_\alpha}{\Box}J+S[\bar \Phi,\Phi]\bigg\}\\
		&\times\exp\bigg\{-mg^2\int d^5z\left(k_\alpha-\frac{i}{2m}\bar D_\alpha J\right)\bigg(\frac{1}{2m}k^\alpha-\frac{i}{2m}\frac{\partial^{\alpha\gamma}}{\Box}\bar D^2\bar k_\gamma+\frac{1}{2}\mathcal{O}^{\alpha\beta}k_\beta\\
		&+\frac{i}{2}\mathcal{O}^{\alpha\beta}\frac{\partial^{\phantom{\beta}\gamma}_\beta}{\Box}\bar D^2 \bar k_\gamma-\frac{i}{2m}\mathcal{O}^{\alpha\beta}\bar D_\beta J\bigg)\bigg\}\\
		&\times\exp\bigg\{-mg^2\int d^5\bar z\left(\bar k_\alpha+\frac{i}{2m} D_\alpha J\right)\bigg(\frac{1}{2m}\bar k^\alpha-\frac{i}{2m}\frac{\partial^{\alpha\gamma}}{\Box}D^2k_\gamma+\frac{1}{2}\mathcal{O}^{\alpha\beta}\bar k_\beta\\
		&+\frac{i}{2}\mathcal{O}^{\alpha\beta}\frac{\partial^{\phantom{\beta}\gamma}_\beta}{\Box}D^2 k_\gamma+\frac{i}{2m}\mathcal{O}^{\alpha\beta} D_\beta J\bigg)\bigg\}.
	\end{split}
\end{equation}
After rewriting all the integrals as an integral over the full superspace, we  obtain the result
\begin{equation}
	\label{final_SD2}
	N_2\int\mathcal{D}\pi^\alpha\mathcal{D}\bar\pi^\alpha
	\exp\left(S_{SD2}\right)=N_2^\pi\exp\left(S_{eff2}[\bar{\Phi},\Phi]\right),
\end{equation}
where $S_{eff2}$ is a effective matter action defined by
\begin{equation}
	\label{effective_2}
	\begin{split}
		S_{eff2}&:=S[\bar \Phi,\Phi]+\frac{g^2}{2}\int 	d^7z\Big[J\mathcal{O}\left(1-\frac{m}{\Box}\bar D^\alpha D_\alpha\right)J-2iJ\mathcal{O}\left(m\frac{i\partial^{\alpha\beta}}{\Box}+C^{\beta\alpha}\right)D_\alpha k_\beta\\
		&+2iJ\mathcal{O}\left(m\frac{i\partial^{\alpha\beta}}{\Box}+C^{\beta\alpha}\right)\bar D_\alpha\bar k_\beta+k^\alpha\left(\frac{2}{\Box}\delta^{\phantom{\alpha}\beta}_\alpha-\mathcal{O}\delta^{\phantom{\alpha}\beta}_\alpha+im\frac{\partial^{\phantom{\alpha}\beta}_\alpha}{\Box}\mathcal{O}\right)D^2k_\beta\\
		&+\bar k^\alpha\left(\frac{2}{\Box}\delta^{\phantom{\alpha}\beta}_\alpha-\mathcal{O}\delta^{\phantom{\alpha}\beta}_\alpha+im\frac{\partial^{\phantom{\alpha}\beta}_\alpha}{\Box}\mathcal{O}\right)\bar D^2\bar k_\beta-2k^\alpha\mathcal{O}\left(i\partial^{\phantom{\alpha}\beta}_\alpha-m\delta^{\phantom{\alpha}\beta}_\alpha\right)\bar k_\beta\Big].
	\end{split}
\end{equation}
It is possible to verify the correctness of this result by varying $S_{eff2}$ with respect to $\Phi$, which leads to the field equation (\ref{finaleqphi}) obtained previously in the classical case. 

The next step is to calculate the functional integral over $V$, as defined in Eq. (\ref{Z2MCS2}). Given that $S_{MCS2}$ is gauge invariant, the differential operator appearing in the quadratic term in $S_{MCS2}$ is not invertible. Consequently, we must fix the gauge using (\ref{GF}), so that the generating functional $S_{MCS2}+S_{GF}$ is given by
\begin{equation}
	\begin{split}
		N_2\int \mathcal{D}V\exp&\left(S_{MCS2}+S_{GF}\right)=N_2\exp\bigg\{\int d^5zg^2k^\alpha k_\alpha+\int d^5\bar zg^2\bar k^\alpha\bar k_\alpha+S[\bar \Phi,\Phi]\bigg\}\\
		&\times\int \mathcal{D}V\exp\bigg\{-\frac{1}{2g^2}\int d^7z V\bigg(\Box\Pi_{\frac{1}{2}}+m\bar D^\alpha D_\alpha+\frac{1}{\xi}\Pi_0\bigg)V\\
		&+\int d^7zV\left(J-iD^\alpha k_\alpha+i\bar D^\alpha\bar k_\alpha\right)\bigg\}.
	\end{split}
\end{equation}
Making use of the result of (\ref{propagator}), we find
\begin{equation}
	\begin{split}
		N_2\int \mathcal{D}V\exp&\left(S_{MCS2}+S_{GF}\right)=N_2^V\exp\bigg\{\int d^5zg^2k^\alpha k_\alpha+\int d^5\bar zg^2\bar k^\alpha\bar k_\alpha+S[\bar \Phi,\Phi]\bigg\}\\
		&\times\exp\bigg\{\frac{g^2}{2}\int d^7z \left(J-iD^\alpha k_\alpha+i\bar D^\alpha\bar k_\alpha\right)\bigg(\mathcal{O}\Pi_{\frac{1}{2}}-\frac{m}{\Box}\mathcal{O}\bar D^\gamma D_\gamma+\xi\Pi_0\bigg)\\
		&\times\left(J-iD^\beta k_\beta+i\bar D^\beta\bar k_\beta\right)\bigg\}.
	\end{split}
\end{equation}
In consequence of the fact that $\Pi_0\left(J-iD^\beta k_\beta+i\bar D^\beta\bar k_\beta\right)=0$, the result gets independent of the gauge-fixing parameter $\xi$. Therefore,
\begin{equation}
	\label{final_MCS2}
	N_2\int \mathcal{D}V\exp\left(S_{MCS2}+S_{GF}\right)=N_2^V\exp\left(S_{eff2}[\bar\Phi,\Phi]\right),
\end{equation}
where $S_{eff2}$ were defined in (\ref{effective_2}).

Note that Eqs. (\ref{final_SD2}) and (\ref{final_MCS2}) show that the generating functional for the spinor self-dual model (\ref{Z2SD2}) and that for the MCS model (\ref{Z2MCS2}) give the same result, up to the field independent factors. This confirms that both models represent the same physical system described by different variables.

\section{Conclusions}

We proved explicitly the duality between $\mathcal{N}=2$, $d=3$ SD and MCS theories both at the classical and quantum levels, for the nontrivial coupling of gauge and SD fields to the external currents. Within our study, we used the master action approach while it is natural to expect that the gauge embedding formalism can be applied as well and yield the same results. Our paper presents two main results. First, it proves this duality in the $\mathcal{N}=2$, $d=3$ superspace for the first time, in contrast to previous studies on the SD and MCS duality within the $\mathcal{N}=1$, $d=3$ superspace \cite{KLRN,FGNPS,FGLNPS}. Second, the equivalence between the models (\ref{MCS2}) and (\ref{SD2}) means that (\ref{MCS2}), which is described by a {\it bosonic} superfield $V(z)$, can be replaced by (\ref{SD2}), which is described by a {\it fermionic} superfield $\pi^\alpha(z)$, without any alteration to the physical content. This result differs from previous studies on the SD and MCS equivalence, which have identified a bosonic/bosonic field correspondence \cite{TPN,DJ,AINRW, GMS} or fermionic/fermionic superfield correspondence \cite{KLRN,FGNPS,FGLNPS}.

Effectively, within our study we demonstrated explicitly, first, the possibility to construct new couplings of $\mathcal{N}=2$ gauge superfields to matter fields. These couplings can be used further to construct new effective theories with $\mathcal{N}=2$ supersymmetry, which, as we expect, can be applied in various contexts including possible extended supersymmetric models for condensed matter, generalizing studies performed in \cite{Abreu}.

An interesting problem consists in possible generalizations of this study. Among various directions of extending our results, it is important to emphasize, first, studying a possibility for their consistent non-Abelian (or noncommutative) generalization, which, as it is known, is still an open question (cf. \cite{Mariz} for a discussion of typical difficulties on this way), second, introducing of other couplings of self-dual/MCS fields to a matter, third, development of a higher-derivative and a non-linear generalization of our models,  generalizing the results obtained in \cite{HD,BI} respectively. We expect to perform these studies in our next papers.

\vspace{5mm}

{\bf Acknowledgments.} The authors thank the Coordena\c{c}\~{a}o de Aperfei\c{c}oamento de Pessoal de N\'{i}vel Superior (CAPES), and the Conselho Nacional de Desenvolvimento Cient\'{i}fico e Tecnol\'{o}gico (CNPq).  Albert Yu. Petrov and  Paulo J. Porf\'irio would like to acknowledge the CNPq, respectively for grant No. 303777/2023-0 and grant No. 307628/2022-1.

\end{document}